\documentclass[twocolumn,prl,aps,superscriptaddress,showpacs,nobalancelastpage]{revtex4}
\usepackage{graphicx}
\usepackage{psfrag}
\usepackage{times}

\def\wf{\omega_f} \def \wc{\omega_c} \def\w{\omega} \def\wa{\epsilon}
\def\stwoa{\sigma^{(2)}_{\alpha \alpha}(0)}
\def\stwo{\sigma^{(2)}_{\alpha \beta}}
\def\szero{\sigma^{(0)}_{\alpha \beta}}
\def\szeroa{\sigma^{(0)}_{\alpha \alpha}(0)}
\def\s{\sigma} \def \sd{\sigma_z}  \def \sx{\sigma_x}
\def\be{\begin{equation}} \def\ee{\end{equation}}
\def\bea{\begin{eqnarray}} \def\eea{\end{eqnarray}}
\def\NL{\nonumber\\}
\def\NLL{\nonumber\\[0ex]}
\def\Sp#1{\mbox{Sp} \left\{\rule[-1ex]{0mm}{2ex}#1\right\}}
\def\va{v_{\alpha}\sigma_-\;} \def\vb{v_{\beta}\sd\;} \def\vx{v_x\sd\;}
\def\vaa{v_{\alpha}}
\def\G#1{G(#1)\;}
\def\dax{\delta_{\alpha\,x}} \def\dbx{\delta_{\beta\,x}}
\def\dab{\delta_{\alpha\,\beta}}
\def\halb{{\textstyle \frac{1}{2}}}
\def\If{\left( \frac{e l_B E_f}{2 \wf} \right)^2}
\def\SS{\scriptstyle}
\def\R{ \Big(
             \begin{array}{cc} \SS 1&\SS 1\\ \SS -1&\SS 1\\ \end{array} \Big) }
\begin{document}

\title{Fractional Minima in the Conductivity of the Quantum--Hall--System
 under Microwaves}

\author{W. Apel}
\affiliation{Physikalisch-Technische Bundesanstalt,
Bundesallee 100, 38116 Braunschweig, Germany.}
\author{Yu.~A. Bychkov}
\affiliation{L.D. Landau Institute for Theoretical Physics, ul.
Kosigina, 2, Moscow, Russia.} \affiliation{Physikalisch-Technische
Bundesanstalt, Bundesallee 100, 38116 Braunschweig, Germany.}
\author{M. Weyrauch}
\affiliation{Physikalisch-Technische Bundesanstalt,
Bundesallee 100, 38116 Braunschweig, Germany.}

\date{\today}

\begin{abstract}
We analyse theoretically the conductivity of a
quantum Hall system exposed to microwave radiation. 
We find that whenever microwave frequency and cyclotron frequency are
commensurate, there is a {\em resonance} in the longitudinal conductivity.
This resonance has the form of the derivative of a Lorentz function;
precisely at the center of the resonance, the microwave induced
conductivity vanishes. Between the resonances there are maxima and
minima, the depths and precise positions of which depend on the
microwave amplitude and the scattering rate of the impurities. 
We demonstrate the existence of these resonances by a microscopic,
analytical calculation of the conductivity in lowest order in the
microwave intensity and show here that the conductivity is independent 
of the microwave polarization, linear or circular. 
We then discuss the general case and predict minima in the longitudinal 
conductivity corresponding to fractional values of the microwave frequency 
divided by the cyclotron frequency.
\end{abstract}


\maketitle

The discovery of new zero--resistance states \cite{MSKNJU02} in the
quantum Hall system under microwave irradiation stimulated a lot of
attention, both experimental,
c.f.~Refs.~\cite{ZDSR01,ZDPW03,D03,Z04,SPSHPW05,M04,SGJPWUDMKK05}
and theoretical, c.f.~Refs.~\cite{DSRG03,AAM03,DVAMP05} and
references therein. As in the original quantum Hall experiment, one
finds regions of vanishing longitudinal resistivity as the magnetic
field varies, but these are not accompanied by plateaux in the Hall
resistivity, instead one sees a classical straight line. Moreover,
the regions of vanishing $\rho_{xx}$ are now labeled by the ratio
$\varepsilon$ of the microwave frequency $\wf$ to the cyclotron
frequency $\wc$, $\varepsilon = \wf / \wc$, while in the case of the
quantum Hall effect, they are classified by the filling factor
$\nu$. There are different theoretical explanations of these
experimental findings. An oscillatory dependence of the conductivity
on $\wf$ was already predicted in early theoretical work by Ryzhii
et al.~\cite{R70,RSS86}. In Ref.~\cite{DSRG03} the
radiation--induced zeros (minima) in the resistivity are seen as the
consequence of photo--exited disorder--scattered electrons
contributing to the resistivity in an oscillatory manner due to the
impurity--broadened energy structure of the Landau levels. On the
other hand, in Refs.~\cite{D03,DVAMP05} an oscillating
non-equilibrium distribution function is invoked. Both quite
different approaches yield negative resistivities in some regions of
$\varepsilon$. Finally, Ref.~\cite{AAM03} complements the picture
with the observation that in regions of negative resistivity a
phase--separation occurs which leads to zero--resistance states. In
spite of all this work, no final consensus about the microscopic
origin of the features occurring in the electronic transport seems
to have emerged.

This situation calls for a careful re--analysis of the microscopic
derivation of the microwave--induced conductivity, in particular
concerning the approximations used. Such an attempt is made in the
present paper. We are lead to a new view on the mechanism
responsible for the structures in the conductivity, as well as the
prediction of minima near fractional values of $\varepsilon$ in very
clean samples. We set ourselves a modest goal: we want to analyze
the \textit{onset} of the effect. Thus, we can expand in the
electric field $E_f$ of the microwave. That, and the use of a simple
model for the impurity scattering makes it possible to carry out an
analytical treatment that enables us to detect the dependence of the
conductivity on its parameters -- up to the final evaluation. In the
first and  major part of this paper we analyze the origin of the
structures in the conductivity up to second order in $E_f$, and at
the end, we make the prediction of minima (zeros) near fractional
values of $\varepsilon$ which appear in higher order in $E_f$. We
extend and modify the work presented in Ref.~\cite{DSRG03} in three
essential ways: First, we do not employ the rotating wave
approximation. That enables us to study arbitrary microwave
polarization, and it will be essential for our prediction. Second,
we couple the microwave field to the velocity, rather than to the
position. We can thus avoid considering vertex corrections in the
case of isotropic impurity scattering. Third, we use a simple model
for the life time of the disordered electronic states. That enables
an analytical treatment.

We consider electrons moving in the x-y plane in a magnetic field
$B\, \hat{e}_z$ (given by its vector potential $\vec A = xB\,
\hat{e}_y$), and in a microwave field with the electric field
polarized in the x-direction (given by $\vec A_f(t) =
\frac{-cE_f}{\wf}\, \sin(\wf t)\, \hat{e}_x$), and in an impurity
potential $V_{imp}(\vec x)$, i.e. the Hamiltonian 
\be
 H = \frac{1}{2m} \left( \vec p - \frac{e}{c} \vec A
                                    - \frac{e}{c} \vec A_f(t) \right)^2
      \;+\; V_{imp}(\vec x) \; .
\ee 
The average over $V_{imp}(\vec x)$ will be discussed below; all
the other symbols not mentioned so far have their standard meaning.
Note that the coupling to the microwave field is different from that
in Ref.~\cite{DSRG03}: 
(i) {\em both} terms $\exp(\mp i\wf t)$, absorption and emission of photons, 
accompany a transition to the next higher Landau level (matrix element of 
$v_x +i v_y$ with $\vec v = (\vec p - \frac{e}{c} \vec A)/m$) and the same 
holds for the Hermitean conjugate process. 
This corresponds to arbitrary linear polarization of the microwave field. 
(ii) the coupling to the microwave field is through the vector potential 
(of the form $v_x (A_{f})_x$). Consequently, we need to include a number of
diamagnetic terms; in this way we can avoid considering vertex
corrections due to the impurities (ladder diagrams) as is discussed
below.

The quantity we want to calculate is the conductivity $\s$. Since
the microwaves introduce a time dependent term into the Hamiltonian,
we will get currents depending on time as $\exp(-i(j\,\wf\!+\!\w
)t)$ for the external electric field $\vec E_{ext}\propto \exp(-i\w
t)$ for any integer $j$. However, in the experiment, one measures
the d.c.~current ($j\,\wf\!+\!\w \to 0$). Therefore, we must neglect
all terms in the current which oscillate in time. In perturbation
theory in $E_f$, we must keep only contributions with the same
number of incoming ($\exp(-i\wf t)$) and outgoing ($\exp(+i\wf t)$)
photon lines. Since we have a non--equilibrium problem, we employ
the Keldysh technique \cite{K65} and use the notations of
Ref.~\cite{Landau-Lifshitz-V10P91}. We apply the standard method of
linear response under an external electric field produced by the
vector potential $c/(i\w)\vec E_{ext} e^{-i\w t}$ and write the
conductivity as
 $\sigma_{\alpha \beta}(\w) = \szero(\w) + \stwo(\w) + {\cal O}(E_f^4) $
where $\szero$ is the conductivity without microwaves and $\stwo$ is
quadratic in $E_f$. Due to the $A_f^2$--term in $H$, we get a number
of ``diamagnetic terms''. We order the result for $\stwo$ w.r.~to
the number of Green's functions $G$, which are calculated in absence
of the microwave field:
\begin{widetext}   \abovedisplayskip-2ex
\bea
 \stwo(\w) =
  \frac{e^2}{2\pi} \; \If \; \frac{1}{\w} \; \int \!\! \frac{d \wa}{2\pi} &&
    \left[ \Sp{\va \G{\wa} \vx \G{\wa-\wf} \vx \G{\wa} \vb \G{\wa-\w}}
   \right. \NL
 && \left. +\halb \Sp{\va \G{\wa} \vx \G{\wa-\wf} \vb \G{\wa - \w - \wf} \vx
   \G{\wa-\w}} \right] \NLL
   + \frac{e^2}{2\pi} \; \If \; \frac{1}{\w} \; \int \!\! \frac{d \wa}{2\pi} &&
    \left[ \Sp{\va \G{\wa} \vx \G{\wa-\wf} \sd \G{\wa - \w} } \dbx \right. \NL
 && \left. +\Sp{\sigma_-\G{\wa} \vb \G{\wa-\w} \vx \G{\wa - \w -\wf} } \dax
   \right.  \NL
 && \left. +\halb \; \Sp{\sigma_-\G{\wa} \vx \G{\wa-\wf} \vx \G{\wa} } \dab
   \right. \NL
  && \left. +\halb \; \Sp{\va \G{\wa} \sd \G{\wa} \vb \G{\wa - \w} }
  \right] \NLL
   + \frac{e^2}{2\pi} \; \If \; \frac{1}{\w} \; \int \!\! \frac{d \wa}{2\pi} &&
  \halb \left[ \Sp{\sigma_-\G{\wa} \sd \G{\wa-\wf-\w}} \dax \; \dbx
  + \halb \; \Sp{\sigma_-\G{\wa} \sd \G{\wa}} \dab \right] \NL
 &&\NL
 && \hspace{-4.4cm} + \left( \wf \to - \wf \right)
                 ;\;+ \left( \w \to -\w, \;\; \mbox{c.c.}\right)
\eea
\vspace{-5mm}
\end{widetext}
\vspace{-5mm} Here, $G$ is a matrix in the $2\times 2$
Keldysh--space, as well as a matrix in the quantum numbers ($n$ and
$k$) of the states for $V_{imp}=0$. $G$ still contains the impurity
potential $V_{imp}$ and the disorder average is implied. The
Pauli--matrices $\sigma_i$ act in the Keldysh--space, the components
of the current $\vec v = (\vec p - e/c \vec A)/m$ in the space of
the quantum numbers $n$ and $k$. All frequencies, or energies, have
been scaled by $\wc$ and are dimensionless, as is the velocity,
since it is scaled by $1/(m l_B)$ where $l_B=\sqrt{c/|eB|}$ is the
magnetic length. The prefactor of the conductivity is the natural
unit of conductance $e^2/(2\pi\hbar)$ ($\hbar$ is restored;
everywhere else, we use units with $\hbar=1$). The dimensionless
quantity 
\be
  I_f \;=\; \If  \;
\ee 
describes the intensity of the microwaves. Its relation to the
parameter $I$ used in Ref.~\cite{DSRG03} contains the
fine--structure constant and the ratio $\varepsilon = \wf /\wc$, 
$ I_f \;=\;  2\pi \; e^2/(\hbar c) \; \wf/\wc \; I  \;.  $

Now we proceed with the evaluation of $\stwo$.
Using the equation of motion for $G$ (and its adjoint) we find
\be
 G(\w) \vb \!\! G(\w) \;=\; G(\w) i \left[H, x_{\beta} \right] \sd G(\w)
\;=\; i \left[G(\w), x_{\beta} \right] \,.
\ee
Using this relation
and $G^{\dagger}(\w) = - \sx G(\w) \sx$, one can convince oneself
that there is no $1/\w$--term in $\stwo$ as $\w \to 0$. This is a
consequence of gauge invariance. We perform the d.c.--limit by
taking the derivative w.r. to $\w$ and specialize to the case of the
diagonal conductivity, $\beta=\alpha$. We now perform the trace in
Keldysh space: after a rotation of the Keldysh matrices (see
Ref.~\cite{Landau-Lifshitz-V10P91})  
\be
  G \;=\; \frac{1}{2}\R \left( \begin{array}{cc} 0&G_A\\G_R&F\\ 
          \end{array}\right) \R^T
\ee 
everything depends on  the retarded and advanced Greens
functions $G_{R}$ and $G_{A}$ only, since the Keldysh function $F$
(related to the ``lesser'' Green's function $G^<$) is given by 
$
 F(\w) = \left[1-2f(\w)\right] \; \left[G_R(\w) - G_A(\w)\right] 
$
where $f(\w)$ denotes the Fermi function.

Now we want to discuss the impurity average.
Before the average, the Green's functions have the form
\be
 G_R(\w) \;=\; \frac{1}{\w - H_0-V_{imp} + i0} \;.
\ee 
For a particle in two dimensions and a transverse magnetic field
there is no dominating class of impurity diagrams that one could sum
up to obtain the self energy. Therefore, we take as a
phenomenological model for the impurity averaged Green's functions
the expression 
\be
 G_R(\w) \;=\; \frac{1}{\w - H_0 + i/({2\tau})} \;.
\ee 
Here, the lifetime $\tau$ enters as a phenomenological parameter
that we fit to the level broadening in the absence of microwaves,
$E_f =0$. We do not intend to calculate $\tau$ from a microscopic
impurity model. Note that $\tau$ does not depend on the Landau
level; as it will turn out, only Landau levels near the Fermi
surface will enter, and as the chemical potential $\mu \approx 60\, \wf$, 
we can safely drop this dependence. Note also that  the
impurity averaged Green's functions become diagonal in the quantum
numbers $n$ and $k$ and independent of $k$ (translational
invariance).

Next we discuss vertex corrections. For the ``non--crossing approximation'' 
in the disorder average~\cite{Abrikosov-Gorkov-Dzyaloshinskii} it is 
well--known that one needs to consider the ``ladder''-type diagrams in the
conductivity in order to have a conserved approximation. But for an
isotropic scattering model, these ladders do not contribute at
\textit{current} vertices. So we take isotropic scattering, and that
enables us to disregard vertex corrections. Note that this would
fail at vertices with \textit{position} operators.

At this point, we give an estimate of the maximum microwave
intensity for which perturbation theory in lowest order is still
applicable. The level distance is $\wc$, the transition matrix
element $ \sqrt{n} \wc \frac{eE_fl_B}{2\wf} $, hence $n I_f^{\rm
max} < 1$. Here, $n$ is the number of the Landau level at the Fermi
energy. There is a more stringent criterion: Calculating the change
of the imaginary part of the self energy in second order
perturbation theory in $E_f$ and demanding that it be smaller than
$1/\tau$, we get 
\be
  n \; I_f^{\rm max} \; \frac{\wc^2}{(\wf-\wc)^2+(\frac{1}{2\tau})^2}
      < 1 \;, \mbox{\hspace{5mm}} (\wf \approx \wc)\;. \label{conditio}
\ee 
In the worst case, at the resonance $\wf=\wc$, we have $n \;
I_f^{\rm max} < (2\wc \tau)^{-2} $, much stronger than the criterion
above.

\begin{figure}[b]
\begin{minipage}{.4\columnwidth}
\psfrag{omega1}[r][r]{$\wa\; n$}
\psfrag{omega2}[l][l]{$\wa\; n_2$}
\psfrag{omega3}[c][l]{$\wa \!-\! \wf \; n_1$}
\psfrag{omega4}[c][l]{$\wa \; n_3$}
\psfrag{alpha}[][]{$\alpha$}
\psfrag{beta}[][]{$\alpha$}
\includegraphics[width=3.5cm]{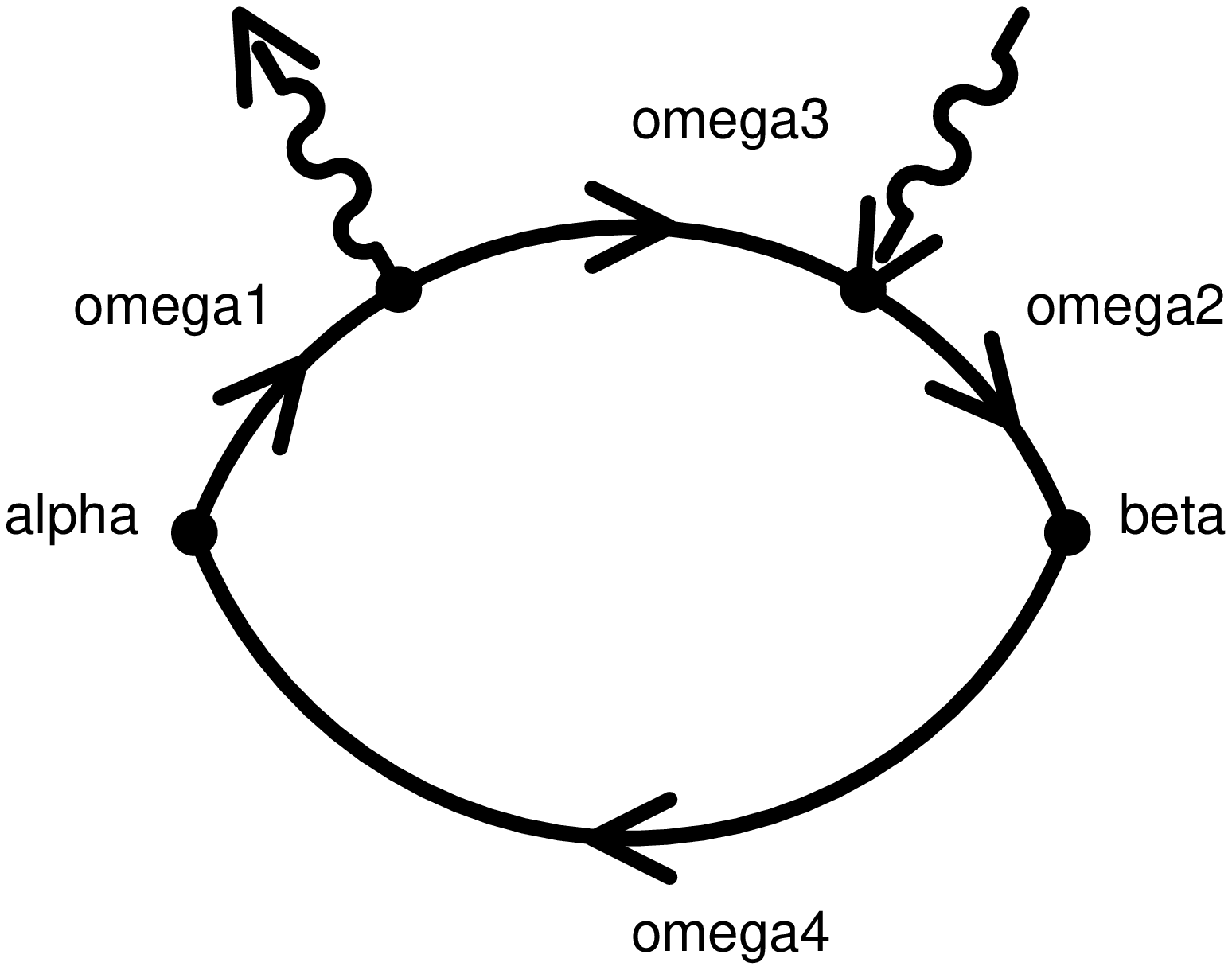}
\end{minipage}\hspace{5mm}
\begin{minipage}{.4\columnwidth}
\psfrag{omega1}[r][l]{$\wa\; n$}
\psfrag{omega4}[l][l]{$\wa \!-\! \wf \; n_2$}
\psfrag{omega2}[l][l]{$\wa \!-\! \wf \; n_1$}
\psfrag{omega3}[r][l]{$\wa \; n_3$}
\psfrag{alpha}[][]{$\alpha$}
\psfrag{beta}[][]{$\alpha$}
\includegraphics[width=3.5cm]{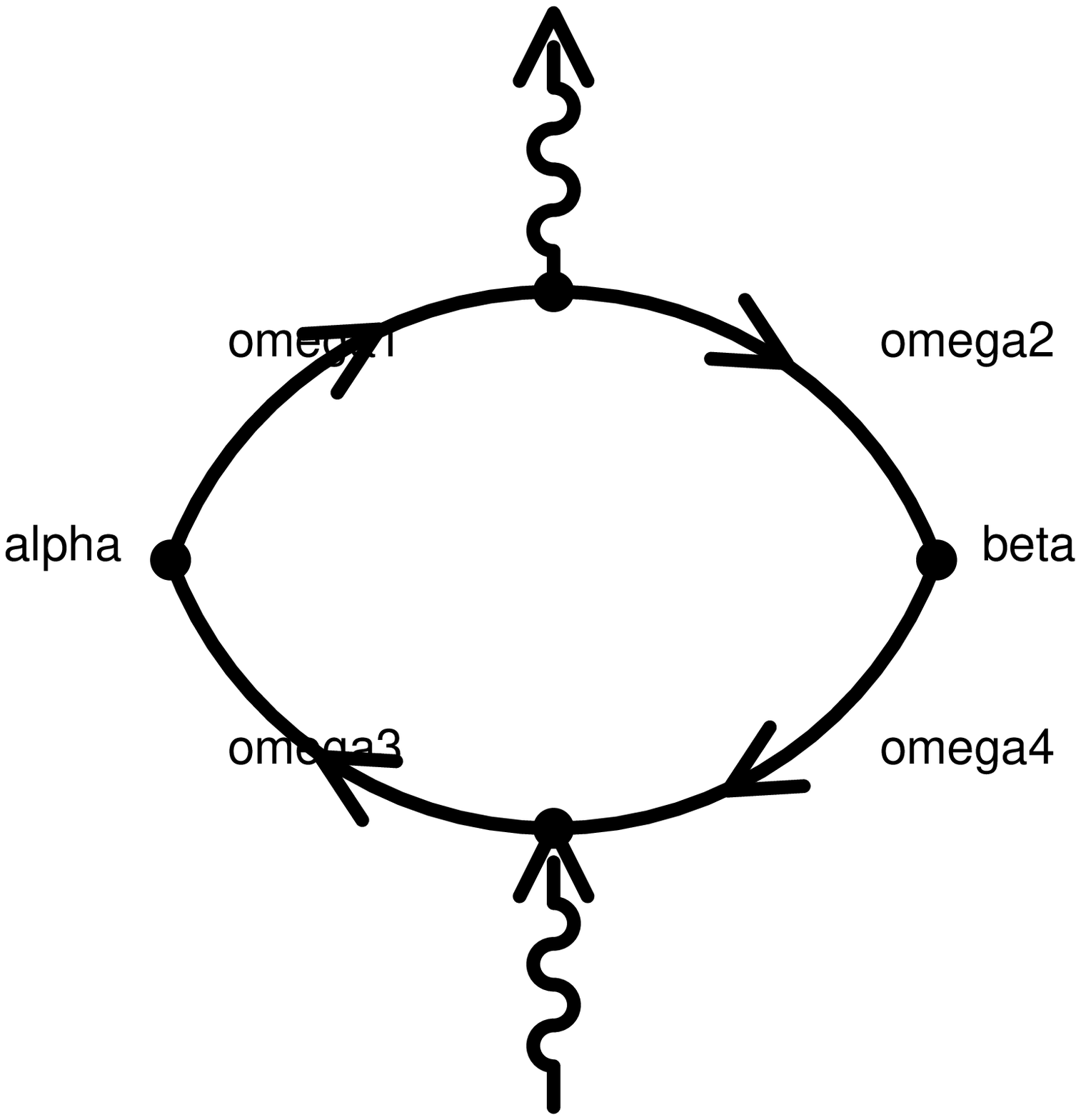}
\end{minipage}
\caption{Diagrams corresponding to the terms with four Green's
functions in Eq.~(\ref{cond})}. \label{diagrams}
\end{figure}
We want to evaluate the formula for the conductivity in the
parameter regime
\be
   1/\tau < \wc \approx \wf < T < \mu   \label{regime}
\ee 
that corresponds to the experimental conditions; $T$ is the
temperature and $\mu \approx 60 \, \wf$. The evaluation of the
trace in the Keldysh space  then leads to expressions of the form
$\int d\wa [f(\wa-\wf) - f(\wa)]$ (or $\wf \to \w$). Thus, strictly
speaking, the conductivity is {\em not} a Fermi surface quantity;
and this is rather obvious in the presence of microwaves. But since
$\wf \ll \mu$, only states close to the Fermi surface enter and we
can approximate $[f(\wa-\wf) - f(\wa)] \simeq -\wf f^{\prime}(\wa)$.
We thus get 
\bea
 && \stwoa \simeq \frac{e^2}{2\pi} \; I_f \; \int \frac{d\wa}{2\pi} \;
 (-) f^{\prime}(\wa) \NL
 && \Big[ -\Sp{\vaa G_R v_x G_R(\wa-\wf)
                 v_x G_R \vaa [G_R - G_A]}  \NL
 && \left. +\wf \Sp{\vaa G_R v_x [G_R-G_A](\wa-\wf)
                 v_x G_A \vaa \partial_{\wa} G_A} \right. \NL
 && \left. -\halb \Sp{\vaa G_R v_x G_R(\wa-\wf) \vaa G_R(\wa-\wf)
                   v_x G_R } \right. \NL
 && \left. +\halb \Sp{\vaa G_R(\wa-\wf) v_x G_R \vaa G_A
                   v_x G_A(\wa-\wf) } \right. \NL
 && \left. -\wf \Sp{\vaa G_R v_x G_A(\wa-\wf) \vaa
                     \partial_{\wa} [G_A(\wa-\wf) v_x G_A]} \right. \NL
 && \left. -\halb \Sp{\vaa G_R G_R \vaa [G_R-G_A] } \right. \NL
 &&   + \left( \wf \to - \wf \right)
                 ;\;+ \left( \mbox{c.c.}\right) \Big].   \label{cond}
\eea 
Here, we omit the argument $\wa$ at the Green's functions
$G_{R,A} \equiv G_{R,A}(\wa)$ for the sake of brevity. In Eq. (\ref{cond}), 
we have dropped several expressions of the type
$[G_R-G_A](\wa) G_R(\wa\!-\!\wf) + \left( \wf \to - \wf \right)$
with two Green's functions with different energy dependence but
without current operator in between. Such terms vanish before the
impurity average ($\propto \delta(\wf)$).

\begin{figure}[b]
\unitlength1cm
\begin{picture}(8,4.8)(0,0)
\psfrag{epsilon}[cb][ct]{\Huge  $\wc/\wf = 1/\varepsilon$}
\psfrag{s}[][lt]{\Huge \
  $s\left(\wf/\wc, 1/(\tau\wc)\right)$}
\put(0,5){\includegraphics[angle=-90,width=8cm]{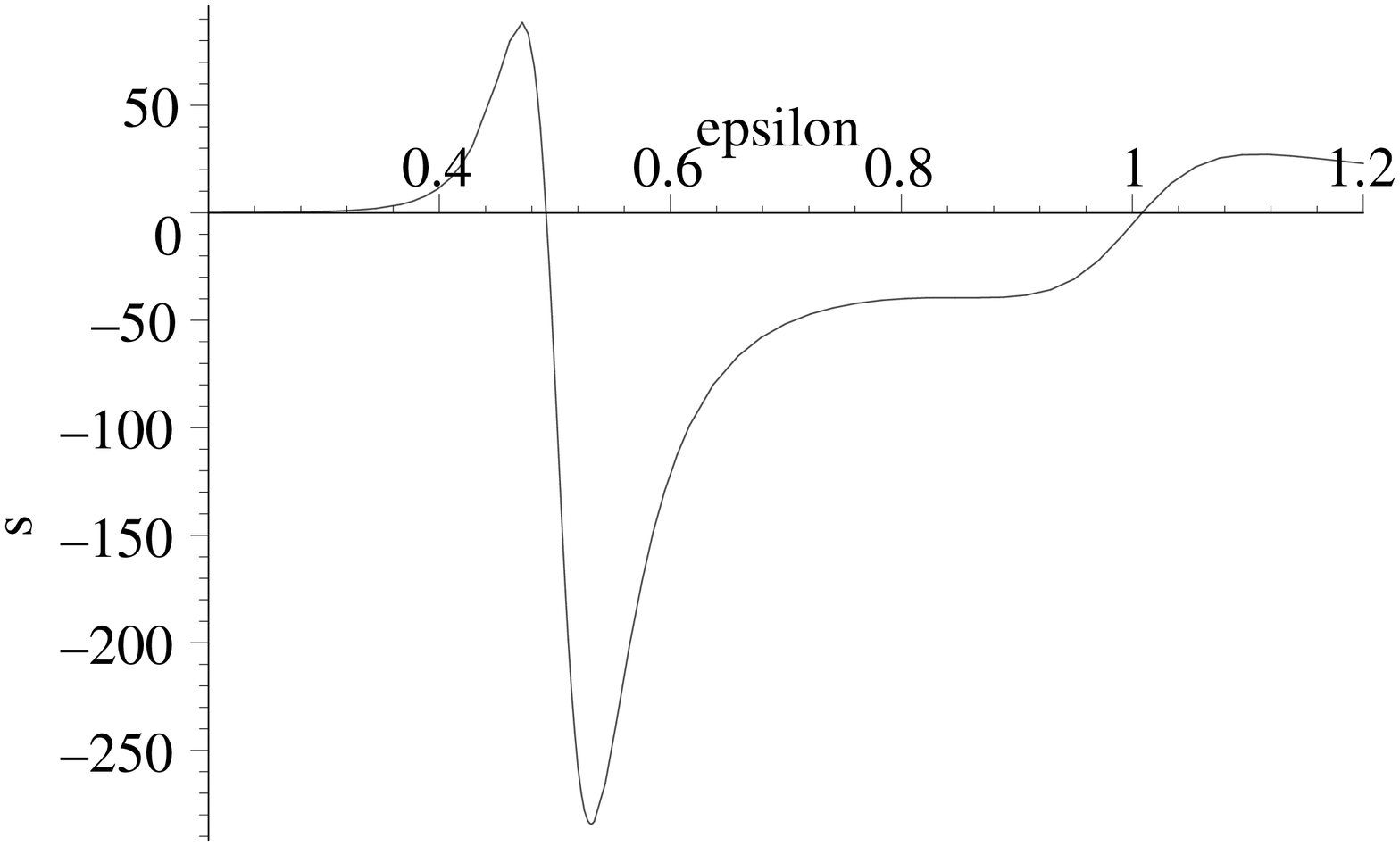}}
\put(4.3,2.8){\includegraphics[angle=-90,width=3.5cm]{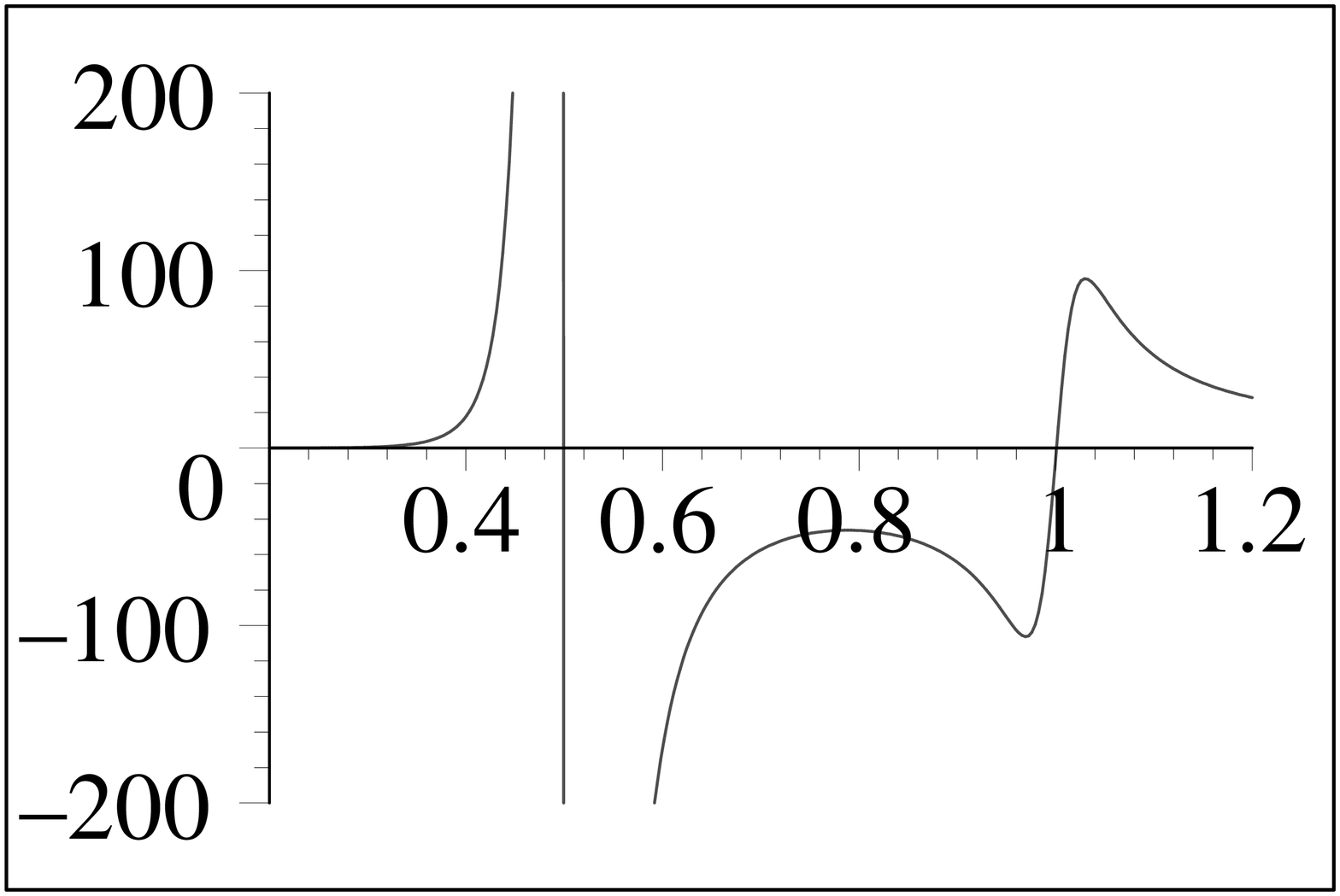}}
\end{picture}
\caption{Contribution of the microwaves to the conductivity relative
to the case without microwaves calculated for $\wf\tau=10$ as a
function of $1/\varepsilon$. Inset: the same calculation for
$\wf\tau=30$. \label{s}}
\end{figure}

It remains to evaluate the integral over $\varepsilon$ and the trace
over the Landau level quantum numbers in Eq. (\ref{cond}). We denote
these quantum numbers  by $n$, $n_1$, $n_2$, and $n_3$ (see
Fig.~\ref{diagrams}). Then, due to the current vertices between the
Green's functions, there are only six combinations $\mathcal C$
($n_1=n\pm 1$, etc.) for $n_1$, $n_2$, and $n_3$ which contribute to
the trace for a given $n$. The integral over $\varepsilon$ and the
sum on $n$ can be evaluated numerically for arbitrary temperature.
For $T=0$, we see strong Shubnikov-de Haas oscillations in $\stwo$.
As we are interested in the regime $\wc<T$, we use the Poisson
summation formula and keep only its first term. Technically, that
corresponds to replacing the sum over $n$ by an integration over
$n$. The lower limit of the integration  is moved from 0 to
$-\infty$, since the main contribution to the integral comes from $n
\sim \mu/\wf$; that is correct in leading order in $\mu/\wf \to \infty$.
The indefinite integral over n is then easily performed by contour
integration and the SdH oscillations disappear.
Finally, we have to perform the sum over the six possibilities
$\mathcal C$ of the quantum numbers $n_i$. This requires some care:
The expression corresponding to the left diagram in
Fig.~\ref{diagrams} for $n_2=n$ contains a self energy insertion,
which must be summed up. It then yields a correction to the single
particle energy as well as to the life time. In the parameter range
considered here such corrections are small and will be neglected.
We write the conductivity as
\be
 \szeroa + \stwoa \simeq \szeroa \left[ 1 + I_f \, \frac{\mu}{\wc} \;
   s\left(\frac{\wf}{\wc}, \frac{1}{\wc\tau}\right) \right] \;
\ee 
Note that the matrix elements $(\vaa)_{n, n+1}\propto \sqrt{n}$
always appear in two pairs in Eq.~(\ref{cond}), and therefore, the
maximum power of $n$ in the numerator is $n^2$, i.e., after the
contour integration, the maximum power of $\mu$ will be $\mu^1$
(since $\szeroa \propto \mu$). It can be shown that there is no
linear order in $\mu$ in $\stwoa$ and we omit lower orders for
$\mu/\wf \sim 60 \gg 1$. The evaluation of the function
$s(\varepsilon, \delta)$ is straightforward; $s$ has the form 
$ s(\varepsilon, \delta) = p(\varepsilon, \delta) / q(\varepsilon, \delta) $
with  {\arraycolsep0pt  \jot1ex
\bea q(\varepsilon, \delta) =
 &&\left[ (2+\varepsilon)^{2}+{\delta}^{2} \right]^{2}
 \left[ (2-\varepsilon)^{2}+{\delta}^{2} \right]^{2}
 \left( \varepsilon^{2}+ \delta^{2} \right)^{2} \NL
 && \left( 1+\delta^{2} \right)
 \left[ (1+\varepsilon)^{2}+\delta^{2} \right]
 \left[ (1-\varepsilon)^{2}+\delta^{2} \right].  \label{q}
\eea }%
The polynomial $p(\varepsilon, \delta)$  is too long to be
printed here and is of maximum order $16$ in $\varepsilon$ and
$\delta$. It vanishes for $\varepsilon \approx \pm 2$ and
$\varepsilon \approx \pm 1$. For $\delta =1/(\wc \tau) = 0.1$, we
get the picture shown in Fig.~\ref{s}.
The conductivity is independent of the relative orientation of the
polarization of the microwaves~\cite{SGJPWUDMKK05} and the
d.c.~current measuring the conductivity. Between $\varepsilon
\approx 2$ and $\varepsilon \approx 1$, the contribution of the
microwaves is negative; thus increasing intensity $I_f$ of the
microwaves diminishes the conductivity until $\sigma$ becomes zero
and one reaches the dissipationless state. Of course, that is
outside the scope of our expansion in $I_f$. Near $\varepsilon = 2$
and $\varepsilon = 1$, there are two points where $\s$ is not
modified by the microwaves. The precise form of the conductivity
depends on $\tau$ (see inset of Fig.~\ref{s}). From Eq.~(\ref{q}) it
is now obvious that the structures in $\s$ are due to {\em
resonances} at $\varepsilon = \pm 1, \pm 2$. (The behavior at
$\varepsilon =0$ is modified by the $\wf$--dependence of $I_f$.)

After having studied the first order approximation in $I_f$, we
finally analyze the resonances in the $N^{th}$ order in $I_f$.
Vertices (including the vertices corresponding to the external
current/vector potential) and lines in a diagram are numbered from
$1$ to $2N+2$ following the arrows. Then, the real part of the
denominator of the $k^{th}$ Green's function is 
\be
  \epsilon_k = \epsilon - \sum_{j=1}^k (\wf \sigma_j + \wc s_j )  \;.
\ee
Here, $s_j = \pm 1$ and $\s_j = 0, \pm 1$ ($\s_j=0$ at the vertices
corresponding to the external current/vector potential) and
$\sum_{j=1}^{2N+2} s_j = 0\;,\; \sum_{j=1}^{2N+2} \s_j = 0$.
A resonance is to be expected whenever $\epsilon_k = \epsilon_{k'}$.
That can occur at $ \varepsilon = \pm p / q $ 
with $p = 1, \cdots, N+1$ and $q = 1, \cdots, N$. In lowest order,
$N=1$, we find $\varepsilon = 1, 2$; in next order $N=2$,
$\varepsilon = 1, 2, 3, 1/2, 3/2$. Note that there is indeed a small
structure (admittedly very small) at $1/\varepsilon = 2/3$ in the
resistivity at 57 GHz in Fig.~1 of the experiment in
Ref.~\cite{ZDPW03}. The order of magnitude of the resonance in $\s$
depends on the degree in the denominator $q$; it should be at least
$I_f^N \wc \tau$.

In this Letter, we have reconsidered the calculation of the
conductivity of a quantum Hall system under microwave irradiation.
In an expansion in $E_f$, the amplitude of the microwave field, we
analyze the form of the non--linear conductivity analytically. We
find that to lowest order the conductivity is independent of the
microwave polarization and determine the maximum intensity for which
perturbation theory is applicable. We see that the structures in the
conductivity are due to resonant scattering of the electrons by the
microwaves. We predict minima (or zeros) in the conductivity near
arbitrary fractions of the parameter $\varepsilon = \wf / \wc$. Note
that our result, a series of resonances with a width determined by
the single particle lifetime $\tau$, is qualitatively different from
a damped sinusoidal behavior. In order to verify this picture one
should do a systematic study of the photoconductivity with
increasing purity of the samples. Then, the observed structure
should get sharper. As the intensity of the microwaves increases,
more and more minima (zeros) should become visible.

\begin{acknowledgments}
One of us (Yu.~A.~B.) thanks the PTB for hospitality and the
Deutsche Forschungsgemeinschaft for financial support (436 RUS
17/21/05). He also acknowledges support of the RFFI 03-02-16012
project. We would like to acknowledge stimulating discussions with
S.~Girvin, J.~Smet, and S.~Dorozhkin.
\end{acknowledgments}

\end{document}